\documentclass[%
 reprint,
 amsmath,amssymb,
 aps,
]{revtex4-1}

\usepackage{graphicx}
\usepackage{dcolumn}
\usepackage{bm}

\usepackage{color,soul}
\soulregister\cite7
\soulregister\ref7

\begin{document}

\title{Enhancement of bulk photovoltaic effect in topological insulators}

\author{Liang Z. Tan}
 \affiliation{The Makineni Theoretical Laboratories,
 Department of Chemistry, University of Pennsylvania, Philadelphia,
 Pennsylvania  19104-6323, USA}
\author{Andrew M. Rappe}
 \affiliation{The Makineni Theoretical Laboratories,
 Department of Chemistry, University of Pennsylvania, Philadelphia,
 Pennsylvania  19104-6323, USA}
 \email{rappe@sas.upenn.edu}

\begin{abstract}
We investigate the shift current bulk photovoltaic response of materials close to a band inversion topological phase transition. We find that the bulk photocurrent reverses direction across the band inversion transition, and that its magnitude is enhanced in the vicinity of the phase transition. These results are demonstrated with first principles DFT calculations of BiTeI and CsPbI$_3$ under hydrostatic pressure, and explained with an analytical model, suggesting that this phenomenon remains robust across disparate material systems.    
\end{abstract}

\pacs{Valid PACS appear here}
\maketitle

The field of topological insulators \cite{moore_birth_2010,hasan_colloquium:_2010} has proven to be a fruitful avenue of research in condensed matter physics in recent years. The symmetry protected surface states of topological insulators \cite{hsieh_observation_2009,chen_experimental_2009,zhang_topological_2009} give rise to novel and useful phenomena such as dissipationless transport \cite{murakami_spin-hall_2004,roushan_topological_2009}. Even though band topology is a ground state property, defined on the occupied valence bands of an insulator\cite{fu_time_2006}, the excited state properties should be affected by topological considerations as well. There is a growing body of research investigating the optical properties of topological insulators\cite{zhong_optical_2015} using photoemission spectroscopy\cite{hsieh_observation_2009,chen_experimental_2009}, Raman spectroscopy \cite{gnezdilov_helical_2011,zhang_raman_2011,zhang_quintuple-layer_2009,bera_sharp_2013}, nonlinear optics \cite{hsieh_nonlinear_2011,hsieh_selective_2011,mciver_theoretical_2012,chen_broadband_2014,muniz_coherent_2014,morimoto_topological_2015}, and photocurrent generation \cite{mciver_control_2012,junck_photocurrent_2013,hosur_circular_2011}. 

While the presence or absence of topological surface states has thus far being the main experimental route for detecting topological phase transitions, in this work we propose an optical, non-surface approach for detecting topological phase transitions in the bulk. This is a direct approach which does not rely on the quality or ease of detection of surface states. We consider the influence of band topology on the bulk photovoltaic effect (BPVE, also known as the photogalvanic effect) \cite{belinicher_photogalvanic_1980,von_baltz_theory_1981}, which is the generation of photocurrents in the bulk of a single-phase material.

The bulk photovoltaic effect is a second-order optical effect which gives current densities, $J_r = \sigma_{rst} E_s E_t$, as a quadratic function of the applied electric fields. This dictates that the BPVE can only be observed in materials with broken inversion symmetry. Measurements of appreciable BPVE have been reported for many ferroelectric materials \cite{chen_optically_1969,glass_high-voltage_1974,dalba_giant_1995,ichiki_photovoltaic_2005}. In the prototypical ferroelectric BaTiO$_3$, experiments \cite{koch_bulk_1975,koch_anomalous_1976} and first-principles calculations \cite{young_first_2012} show that the primary mechanism for BPVE is the shift current \cite{sipe_second-order_2000}. In this mechanism, carriers are excited into a current-carrying superposition of excited states. Due to its dominant effect in the bulk photovoltaic response, we will focus on the shift current in this paper.

Because optical excitations probe both the valence and conduction band wavefunctions, we expect the band inversion process, which is the interchange of the conduction and valence band characters, to result in dramatic changes in the finite-frequency response functions of the system. We show that band inversion reverses the direction of the shift current. The magnitude of the shift current is enhanced in the vicinity of the band inversion transition. We illustrate these results using first-principles density functional theory (DFT) calculations of the polar topological insulators BiTeI and CsPbI$_3$. Under hydrostatic pressure, these materials undergo band inversions and enter topologically non-trivial phases, where the magnitude of shift current is enhanced to an order of magnitude larger than that of BaTiO$_3$. 

\begin{figure}
\includegraphics[width=0.49\textwidth]{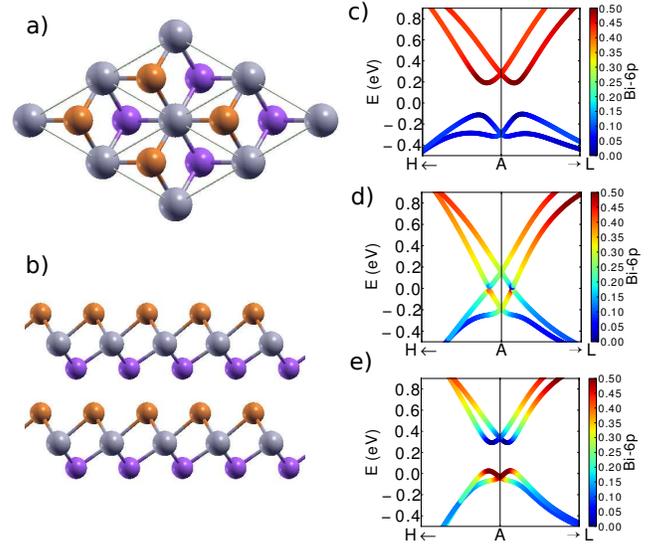}
\caption{Structure of BiTeI displayed in top view (a), and side view (b). Grey: Bi, purple: Te, orange: I. Panels (c) - (e): Evolution of band structure across the topological phase transition, from the normal insulator phase (lattice constant $a=4.42$ \AA) (c), to the band inversion transition ($a=4.18$~\AA) (d), to the topological insulator phase ($a=4.05$ \AA) (e). Colors represent the projection of the wavefunctions to Bi 6$p$ orbitals. The band character changes rapidly at the band inversion transition (d).}
\label{fig1}
\end{figure}

Being a second-order response function, the shift current bulk photovoltaic response function includes virtual transitions of electrons to all unoccupied bands. 

\begin{equation} 
\label{ppp}
\begin{aligned}
\sigma_{rst}(\omega) = & \frac{e}{m} \left (\frac{e}{m\hbar\omega} \right )^2 \sum_{123} [f(1)-f(2)] \delta(\pm \omega-\omega(1)+\omega(2)) \\
\times & \Bigg [ 
\langle 1 \lvert p_r \rvert 2 \rangle 
\langle 2 \lvert p_s \rvert 3 \rangle
\langle 3 \lvert p_t \rvert 1 \rangle
\mathcal{P}\frac{1}{\omega(3)-\omega(1)}\\
& -(1\rightarrow3, \,2\rightarrow1,\,3\rightarrow2) \Bigg ]
\end{aligned}
\end{equation}

\noindent
Unlike linear response (e.g., Fermi's golden rule), interchanging the conduction band and valence band in Eq.~\ref{ppp} changes the value of the response function. In Eq.~\ref{ppp}, the sum over states includes all conduction and valence bands and corresponding integrals over the Brillouin zone. The components of the momentum operator are denoted by $p_r$. The shift current can be re-expressed as a product of the absorption rate and the shift vector $R$.

\begin{equation}
\label{sc}
\begin{aligned}
\sigma_{rst}(\omega) = \pi e  \left (\frac{e}{m\hbar\omega} \right )^2
\sum_{cvk} &
\langle c \lvert p_r \rvert v \rangle 
\langle v \lvert p_s \rvert c \rangle \\
&\quad\delta(\omega_c - \omega_v - \omega) R_t(c,v,k)
\end{aligned}
\end{equation}

\begin{equation}
\label{sv}
R_t(c,v,k) = -\frac{\partial}{\partial k_t} \arg \langle c \lvert p_r \rvert v \rangle - [\chi_{vt}(k) - \chi_{ct}(k)]
\end{equation}

\noindent
Here, $\chi$ are Berry connections. The shift vector $R$ can be understood as a generalized gauge invariant $k$-space derivative of the $p$ operator \cite{aversa_nonlinear_1995,sipe_second-order_2000}, and it is odd under the interchange of $c$ and $v$ bands.

We apply this formalism to perform first-principles DFT calculations \cite{young_first-principles_2012,brehm_first-principles_2014,zheng_first-principles_2015,wang_first-principles_2015} of the shift current BPVE of the polar layered material BiTeI. This material is a normal insulator under ambient pressure, and DFT predicts that it becomes a topological insulator at pressures higher than 1.7 GPa \cite{bahramy_emergence_2012}. It lacks inversion symmetry on either side of the topological phase transition. The strong spin-orbit interaction, together with the breaking of inversion symmetry, result in a Rashba-type splitting \cite{das_engineering_2013,picozzi_ferroelectric_2014,zhong_giant_2015} of the conduction and valence bands in both topological insulator (TI) and normal insulator (NI) phases. At the topological phase transition, the band gap closes between the lowest energy Rashba conduction band and the highest energy Rashba valence band, in the vicinity of the $A$ point in the Brillouin zone (Fig.~\ref{fig1}). Due to trigonal warping effects, the band gap closes at six symmetry-related points along the $A-H$ directions. Nevertheless, because the magnitude of trigonal warping is small, there is an approximate ring-like degeneracy of the conduction and valence bands at the topological phase transition at an energy scale above 10 meV. In general, for topological phase transitions in three dimensions, it is expected that the band gap closes over a range of pressures \cite{murakami_phase_2007}. However, this pressure range is very small in BiTeI, appearing only at higher orders in $k$ \cite{bahramy_emergence_2012}, and does not affect the nature of our results presented below. Electronic correlation effects beyond DFT are likely change the transition pressure, but not the presence of a band inversion transition \cite{jin_topological_2012}.

We have calculated band energies and electronic wavefunctions within DFT, using the PBE exchange-correlation energy functional and norm conserving pseudopotentials. A 8$\times$8$\times$8 Monkhorst-Pack $k$-grid and plane wave energy cutoffs of 50 Ry were used for self-consistent evaluation of the charge densities. A refined $k$-point sampling near the A points equivalent to a 96$\times$96$\times$96 mesh revealed changes in the shift current spectrum as the system is tuned across the topological phase transition. For simplicity, we restrict our attention to the $\sigma_{zzz}$ tensor component of the bulk photovoltaic response in this paper. 

\begin{figure}
\centering
\includegraphics[width=0.49\textwidth]{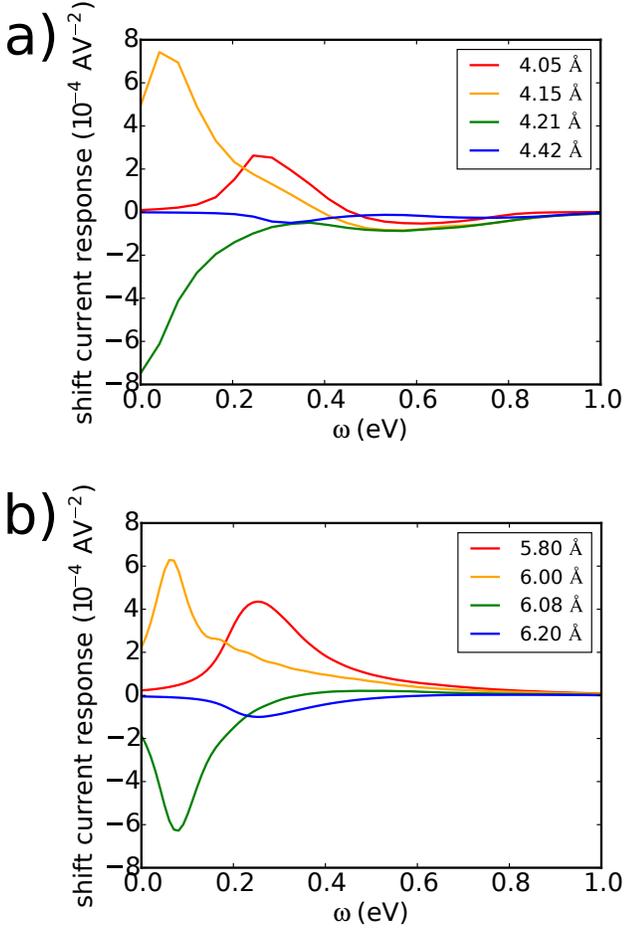}
\caption{Shift current bulk photovoltaic response of BiTeI (a) and CsPbI$_3$ (b). In each panel, spectra are given for a range of lattice constants, specified in \AA. }
\label{fig2}
\end{figure}

\begin{figure}
\centering
\includegraphics[width=0.49\textwidth]{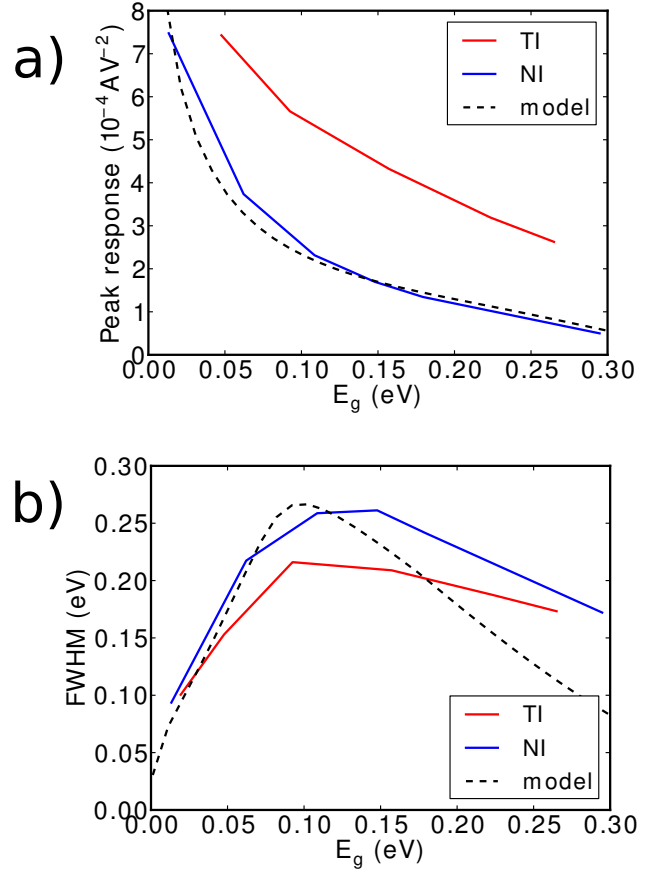}
\caption{Spectral parameters of the shift current response for BiTeI. Absolute value of the peak shift current is shown in (a) as a function of different band gap energies, corresponding to different applied pressures and lattice constants. Full width at half maximum of the shift current peak is shown in (b) as a function of different band gap energies, corresponding to different lattice constants. Dashed lines show the prediction of the model developed in Eq.~\ref{scdos}, using dipole matrix elements fitted to the normal insulator (NI) phase.}
\label{fig3}
\end{figure}

We focus on the low-energy shift current spectrum, with contributions from the lowest-energy conduction and valence bands. The dominant spectral feature in this energy range is a peak located at the band gap energy (Fig.~\ref{fig2}). Starting in the NI phase, an increase in pressure reduces the band gap, driving the shift current peak to lower energies while simultaneously increasing its magnitude. At the critical pressure for band inversion, the shift current peak abruptly changes sign. Further increasing the pressure increases the band gap, shifting the peak to higher energies and reducing its magnitude. Comparing the magnitude of the shift current peak in TI and NI phases, we have found that the peaks in the TI phase are consistently larger in magnitude than in the NI phase (Fig.~\ref{fig3}). 
The peaks in either phase are sharpened as the band inversion transition is approached. 

In brief, the change in direction of the shift current at the band inversion transition is a result of the shift vector $R$ flipping sign upon interchange of conduction and valence bands. The enhancement of the shift current is a consequence of the rapidly changing band characters at the band inversion transition (see SI-II), as measured by the generalized $k$-derivative in the shift vector Eq.~\ref{sv}. In addition, the smaller bond lengths in the TI phase increase the magnitude of the momentum matrix elements, leading to the enhancement of the shift current in the TI phase over the NI phase.

To further understand these results, we develop a low energy effective model describing the band inversion process. We start with a 4-band Dirac model $H_0$ which has been used to describe topological phase transitions in centrosymmetric systems, such as semiconductor quantum wells \cite{bernevig_quantum_2006} and perovskites \cite{jin_topological_2012}.

\begin{equation}
\label{h0}
H_0(\vec{k}) = M(k)\tau_z \otimes \sigma_0 - i\hbar v_F \tau_x \otimes (\vec{k} \cdot \vec{\sigma})
\end{equation}

\noindent
Here, $\vec{\tau}$ and $\vec{\sigma}$ are pauli matrices representing orbital and spin degrees of freedom respectively. The second term in Eq.~\ref{h0} describes the band structure at the point of band inversion, where it is gapless, and has linearly dispersing bands with band velocity $v_F$. The first term introduces a band gap away from the band inversion, controlled by the mass term $M(k)=m - A k^2$. We have taken the time-reversal operator to be $T=\tau_z \otimes \mathcal{K} i\sigma_y$. As inversion symmetry breaking is a prerequisite for shift current, we add an inversion symmetry breaking term $H_{ISB}$, which preserves the time-reversal symmetry of $H_0$. To lowest order in $k$, all such inversion symmetry breaking terms can be written as $\tau_x \otimes \sigma_i$, which introduces a polar axis in direction $i$. Without loss of generality, we choose 
$H=H_0 + H_{ISB}$, 
$H_{ISB}=\mu \, \tau_x \otimes \sigma_z $. $H_{ISB}$ is responsible for the Rashba split conduction and valence bands (Fig.~\ref{fig1}).



Only the lowest-energy Rashba conduction and valence bands participate in the band inversion transition. As such, we reduce $H$ to a $2\times2$ Hamiltonian $H'$ for these bands by means of a canonical transformation (SI). The reduced Hamiltonian is

\begin{equation}
\label{hp}
\begin{aligned}
H' &= W^{\dagger} H W \\
   &= 
\begin{pmatrix}
M(k) & \lambda \\
\lambda & -M(k)
\end{pmatrix}
\end{aligned}   
\end{equation}

\noindent
with band energies 

\begin{equation}
\label{epm}
E_{\pm}(k) = \pm \sqrt{M(k)^2 + \lambda^2}
\end{equation}

\noindent
where $\lambda = \sqrt{(\hbar v_F k_z)^2 + (\mu -\hbar v_F k_\perp)^2}$. From Eq.~\ref{epm}, we see that the band inversion transition happens when the parameter $m$ is tuned to $m^*= A\mu^2 / \hbar^2 v_F^2$. The band gap closes in the ring $k_z=0, k_{\perp}=\mu / \hbar v_F$, with the size of the ring depending on the amount of inversion symmetry breaking ($\mu$). From the relative signs of the mass term in Eq.~\ref{h0} at $k\approx0$ and $k\rightarrow\infty$, it can be shown that $m>m^*$ corresponds to the inverted band structure, while $m<m^*$ corresponds to the normal, uninverted phase \cite{bernevig_quantum_2006}, when $A>0$. In BiTeI, $m$ can be controlled by applying pressure.

We now proceed to obtain the $\sigma_{zzz}$ tensor component of the shift current for the effective Hamiltonian $H'$. In the basis of $H$, the momentum operator $\vec{p}$ is taken to be $\tau_x \sigma_i$, by the same argument for the form of $H_{ISB}$. The shift current contribution from the low energy bands is obtained by reducing to the basis of $H'$ using the transformation $W^\dagger p W$, and applying Eq.~\ref{sc} (SI). 

\begin{equation}
\label{scdos}
\begin{aligned}
\sigma(\omega) &= \pi e  \left (\frac{e}{m\hbar\omega} \right )^2
\lvert p \vert^2 \,
\textrm{DOS}(\omega) \theta(\omega-E_g) \frac{M}{E_{+}}  \\
&= \pi e  \left (\frac{e}{m\hbar\omega} \right )^2
\lvert p \vert^2 \,
\textrm{DOS}(\omega) \theta(\omega-E_g) \frac{M(\omega)}{\omega/2}
\end{aligned}
\end{equation}

\noindent
Here $\lvert p\rvert^2$ is the dipole matrix element, assumed to be constant in this range of $k$, and $M(\omega)= \left . M(k) \right \rvert_{\omega=E_+(k)-E_-(k)}$. Eq.~\ref{scdos} shows that the shift current lineshape is a peak located at the band gap energy $\omega=E_g$. The direction of shift current is dictated by the sign of $M$. Across the band inversion transition, the mass term $M$ changes sign, as does the shift current response, in agreement with the results of DFT (Fig.~\ref{fig2}). We note that the direction of the shift current in a real material is a consequence of many factors \cite{young_first_2012}; band inversion flips the sign of the shift current, but does not fix a particular direction for the current.

The enhancement of the DFT shift current in the vicinity of the band inversion transition is well explained by this model (Fig.~\ref{fig3}), and is consistent with the correlation between band gap and shift current reported in \cite{cook_design_2015}. Likewise, the sharpening of the shift current peak near the band inversion transition can be understood by writing $M/E_{+} = M/\sqrt{M^2+\lambda^2}$ in Eq.~\ref{scdos}. When the mass term $M$ is small, it determines the width of the peak, as seen from the agreement between the model and DFT in Fig.~\ref{fig3}. As $M$ is increased, it reaches a point where it becomes comparable to the Rashba splitting energy scale, which limits the range of applicability of Eqs.~\ref{hp}-\ref{scdos}.

\begin{figure}
\centering
\includegraphics[width=0.49\textwidth]{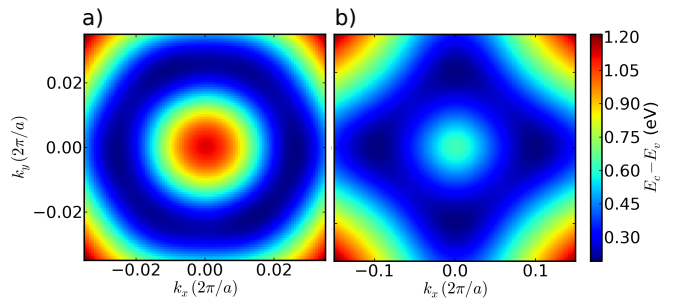}
\caption{Low energy band structure of BiTeI (a) and CsPbI$_3$ (b) near the band inversion transitions. Energy difference between the lowest conduction band and highest valence band is shown at $a=7.90$~\AA\ for BiTeI (a) and at $a=6.075$~\AA\  for CsPbI$_3$ (b).}
\label{fig4}
\end{figure}

We expect the qualitative nature of these results to remain unchanged in the presence of higher-order band warping effects. To demonstrate this, we calculate the shift current response of the halide perovskite CsPbI$_3$ using DFT. Under hydrostatic pressure, CsPbI$_3$ becomes a ferroelectric topological insulator \cite{liu_hydrostatic_2015}. The band structure in the vicinity of the topological phase transition is similar to that of BiTeI, except for a warping of the low energy bands, which breaks the ring-like minima of the Rashba bands into four shallow, degenerate valleys (Fig.~\ref{fig4}). Nevertheless, the shift current response of CsPbI$_3$ shows similar trends as that of BiTeI (Fig.~\ref{fig2}). The main differences arise from the different density of states of BiTeI ($\textrm{DOS}(\omega)\sim \sqrt{\omega-E_g}$) and CsPbI$_3$ ($\textrm{DOS}(\omega)\sim (\omega-E_g)^{3/2}$), which give rise to somewhat different line shapes, and a weaker enhancement of shift current for CsPbI$_3$.

Shift current materials have been suggested as candidates for light-harvesting energy applications, due to active photocurrent generation in the entire bulk of the material and their above-band gap photovoltages \cite{ji_bulk_2010}. The BPVE is a ballistic hot carrier effect. As such, a material with a small band gap can still have carriers emerge with more than the band gap worth of energy, as they do not thermalize before leaving the photovoltaic material. Recent works \cite{bhatnagar_role_2013,wang_first-principles_2015,young_first-principles_2015,wang_substantial_2016} have studied how to achieve a large shift current magnitude in real materials, by considering structural modifications such as domain walls and chemical alloying. Our result suggests a new way to design shift current materials based on their electronic topology: to look for materials close to a band inversion transition. Furthermore, the shift current enhancement reported in this paper may find applications in high-resolution sensors and switches, due to its robustness and the abruptness of the change in current direction. For instance, the direction of the photocurrent can be controlled by small changes in the lattice constant, flipping directions as the material undergoes a band inversion transition. Complex functionality coupling electrical, optical, and mechanical degrees of freedom may be implemented simply in a bulk crystal of these topological shift current materials.

We have demonstrated clear signatures of band inversion in the shift current response of noncentrosymmetric materials. As a material is tuned across a topological phase transition, we predict a reversal of the photocurrent direction, sharpening of the shift current peak, and enhancement of the photocurrent during band inversion and in the inverted phase. These features are robust across various material systems such as the perovskite CsPbI$_3$ and the layered BiTeI. Besides shedding light on the excited state properties of topological insulators and providing a bulk optical signal of band inversion, these results suggest a route for enhancing the bulk photovoltaic effect by designing materials close to band inversion, with possible applications in light-harvesting devices. 

\begin{acknowledgments}

We thank C. L. Kane, E. J. Mele, Y. Kim, and F. Zheng for enlightening discussions.
L.Z.T. was supported by the U.S. Office of Naval Research, under grant N00014-14-1-0761.
A.M.R. was supported by the U.S. Department of Energy, under grant DE-FG02-07ER46431.
The authors acknowledge computational support from the NERSC of the DOE and the HPCMO of the DOD.

\end{acknowledgments}

\bibliographystyle{apsrev}
\bibliography{main}

\end{document}